\documentstyle[manuscript,prb,aps]{revtex}
\newcommand{\bey}[1]{\begin{eqnarray} \label{#1}}
\newcommand{\eey}{\end{eqnarray}}
\newcommand{\beq}[1]{\begin{equation} \label{#1}}
\newcommand{\eeq}{\end{equation}}

\begin{document}
\bibliographystyle{prsty}

\title{Strict Detailed Balance is Unnecessary in Monte Carlo Simulation}
\author{Vasilios I.\ Manousiouthakis  and Michael W.\ Deem\\
Chemical Engineering Department\\
University of California\\
Los Angeles, CA\ \ 90095--1592}
\maketitle
\renewcommand{\baselinestretch}{1.3} \tiny \normalsize
\renewcommand{\arraystretch}{.69}

\begin{center}
{\bf{Abstract}} 
\end{center}
\begin{quote}
Detailed balance is an overly strict condition  to ensure a valid
Monte Carlo simulation. We show that, under fairly general
assumptions, a Monte Carlo simulation need satisfy only the
weaker balance condition.  Not only does our proof show that
sequential updating schemes are correct, but also it establishes the
correctness of a whole class of new methods that
simply leave the Boltzmann distribution invariant.
\end{quote}
\vspace{2in}
To appear in \emph{J.\ Chem.\ Phys.}
\newpage

The theoretical basis for Monte Carlo simulation is
Markov chain theory.  In a Monte Carlo simulation, the
configuration of a system of interest 
is updated by a series of moves, and the moves are chosen so that
the updating process forms a Markov chain.  Let us consider, without
 loss  of generality, that the system is in the canonical ensemble.
We then desire that the configurations generated by the
Markov chain sample the Boltzmann distribution, after an initial
transient, ``equilibration'' period.  We would like, in other
words, that the limiting distribution of the Markov chain exists,
is unique, and is the Boltzmann distribution.  This
result is assured if the Markov chain is \emph{regular} and
satisfies the detailed balance condition.\cite{ParisiBook}
The proof of this result uses
the Perron-Frobenius theorem and the fact that a matrix obeying
detailed balance has a complete set of eigenvectors.

There are many Monte Carlo methods in wide use, however, that
do not satisfy detailed balance.  The classic example is
the  sequential updating method.\cite{FrenkelSmit}
  In such a method, the reverse
move of sequentially updating in the reverse order never occurs,
and so the method cannot satisfy detailed balance.
The correctness of these simulations cannot be justified with
the standard proof, which relies on the detailed balance condition.
In fact, there appears to be no proof available in the literature
that would assure us that these methods are correct.  Yet, they are
in wide use.  Many specific examples can be given.
One recent example is sequential updating of spins in the Ising
model.\cite{marinari98}  Another is sequential updating of
surface patches in membrane simulations.\cite{Marco}
Of interest to the chemistry and chemical engineering community 
is the original implementation of the Gibbs ensemble, where
particle displacements, volume changes, and particle exchanges
were carried out in a fixed order.\cite{Panagiotopoulos}
All implementations to date of parallel tempering, a powerful method for
sampling ``glassy'' systems, perform a fixed sequence
of particle moves and system swaps.\cite{marinari98,geyer91,geyer95,hukushima96,tesi96,tesi96b,Boyd98,hansmann97}
Finally, several
hybrid methods that combine Monte Carlo and molecular dynamics
in a sequential fashion have been introduced.\cite{Chandler}
A variety of more general methods have been introduced that 
are guaranteed only to leave the Boltzmann distribution invariant,
not to converge to it.\cite{Wing}
Given the lack of a rigorous proof of correctness of these
methods and the general  belief
in the importance of detailed balance, these methods have 
been criticized by various authors.

In this short communication, we show that detailed balance is
too strict a condition.  In particular,
we show that the weaker balance condition is sufficient and necessary.
The balance condition simply requires that the Markov chain
leave the Boltzmann distribution invariant.
To be precise, we show 
that a set of Monte Carlo moves is valid if they
1) lead to Markov sampling, 
2) lead to \emph{regular} sampling, and
3) satisfy the balance condition.

The organization of our paper is as follows.  We first formulate
in matrix notation the master equation that governs the Markov process.
Next, we state precisely our theorem.  We then prove
the theorem, first in a heuristic way and then in a
rigorous, formal fashion.
  We conclude with a discussion of the implications of
this theorem.

A Monte Carlo simulation can be understood as
an implementation of a discrete-time master
equation.  The state of the system at
$n$ Monte Carlo steps (MCS) is related to that
at $n-1$ MCS by a transition matrix:
\beq{1}
{\bf x}_n = A {\bf x}_{n-1} \ ,
\eeq
where the probability that the system is in state
$i$ at $n$ MCS is given by the i-th component of the vector
vector ${\bf x}_n$, and $A$ is the transition matrix.
In this formulation, the sequence of steps forms a
Markov chain.  For a method that satisfies detailed balance,
each step can be one of the many possible moves, chosen at
random.  For a sequential updating procedure, each step
will be an entire sweep through a sequential series of
moves.

The transition matrix satisfies certain obvious properties.
The first is that the matrix is non-negative:
\beq{2}
A_{ij} \ge 0   ~~~~~\forall i,~~ \forall j \ ,
\eeq
 which ensures that
the probabilities remain positive.  The second obvious condition
is that  the transpose of the matrix is stochastic:
\beq{3}
\sum_i A_{ij} = 1 ~~~~\forall j \ ,
\eeq
which ensures that the transition matrix conserves total probability.
It is well known that the set of Monte Carlo  moves must lead to
ergodic sampling, otherwise the system will not sample all of
phase space.  In matrix notation, this is expressed as
$[A^m]_{ij}>0$ for all $i$ and $j$, for some $m$, 
which may depend on $i$ and $j$.
 We will need the slightly stronger condition
that the set of moves leads to \emph{regular} sampling.  This
condition is expressed as
\beq{4}
[A^m]_{ij} > 0~~~~~ {\rm for~all~}i,j{\rm~for~some~fixed~}m \ .
\eeq
Note that a set of moves
that leads to ergodic sampling can be converted
into a set of moves that leads to regular sampling with the addition
of some null moves:
\beq{4a}
A' = (1- \alpha) A + \alpha I, ~~~~~ (0 < \alpha < 1) \ ,
\eeq
where $I$ is the identity matrix.
Finally, if the desired distribution, \emph{i.e.}\ the Boltzmann
distribution, is denoted by ${\bf x}^*$, the balance condition is 
simply that 
\beq{5}
A {\bf x}^* = {\bf x}^* \ .
\eeq
We note that the detailed balance condition is expressed in
matrix notation as $A_{ij} x^*_j = A_{ji} x^*_i$.
We further note that if matrix $A$ satisfies the balance condition,
then the matrix $A'$ in Eq.\ (\ref{4a}) also satisfies the
same balance condition.

Our theorem is that a set of Markov moves satisfying conditions
(\ref{2}-\ref{4}) always converges to a limiting distribution, which
is unique.  It follows that if
 by some means we know that the moves satisfy the
balance condition (\ref{5}), then the simulation will eventually 
converge
to sampling the desired distribution ${\bf x}^*$.

We first present a simple proof of this theorem.  The
Perron theorem implies that the matrix $A^m$ has a single,
unique maximum eigenvalue, $\lambda$, with an associated eigenvector,
${\bf x}^*$, with non-negative
 entries (Theorems 1.4.4 and 1.8.1 of Ref.\ \onlinecite{Bapat}).
By taking a sum over components in the equation $A^m {\bf x}^* =
\lambda {\bf x}^*$, we conclude that $\lambda = 1$.
The limiting distribution, if it exists, is given by
$\lim_{n \to \infty} A^{mn} {\bf x}_0$, where 
${\bf x}_0$ is the initial condition.  Since the sampling is
regular, this limit cannot depend on the initial condition, 
because essentially all initial conditions are sampled during the
initial equilibration period.  By choosing an initial condition of
${\bf x}_0 = {\bf x}^*$, we conclude that the limiting distribution
of $A^m$ is ${\bf x}^*$.  Since the initial condition is
irrelevant, we can equally well consider limits
such as $\lim_{n \to \infty} A^{mn} [A^k {\bf x}_0]$, which simply
correspond to different choices for initial conditions.
But these limits all must be ${\bf x}^*$ for any $k$ by the same
argument.  That all of these limits are equal implies that
\beq{6}
\lim_{n \to \infty} A^n {\bf x}_0 = {\bf x}^* \ ,
\eeq
where the $m$ has been eliminated, and the theorem is proved.

We now present an alternative and more formal proof of the theorem.
From conditions (\ref{2}) and
 (\ref{3}) and Lemma 3.1.1 of Ref.\ \onlinecite{Bapat} we conclude
that the spectral radius of $A$ is unity.  From conditions
(\ref{2})-(\ref{4}) and the Perron-Frobenius theorem (Theorem
1.4.4 of Ref.\ \onlinecite{Bapat})
we conclude that 1 is an eigenvalue of $A$ and that the
components of the
corresponding eigenvector are non-negative.  We denote this eigenvector
by ${\bf w}^{(1)}$.  From condition (\ref{4}) we can, in fact,
reach the stronger conclusion that all other eigenvalues of $A$ have
a modulus strictly less than unity (Theorem 1.8.1 of Ref.\
\onlinecite{Bapat}).
Page 40 of Ref.\ \onlinecite{Bapat} then states that
\beq{7}
\lim_{n \to \infty} A^n =  
\lim_{\lambda \to 1^+} 
\left[
(\lambda -1)(\lambda I -A)^{-1} \right] \ .
\eeq

We now proceed to evaluate the limit in Eq.\ (\ref{7}).
We first decompose the $l \times l$ 
matrix $A$ 
into its Jordan normal form:\cite{Grantmacher}
$A = W J W^{-1}$,
 where 
\beq{8}
J = \left[ 
\begin{array}{ccccc}
\lambda_1 & b_1&&&0 \\
& \lambda_2 & b_2\\
&  & \ddots & \ddots \\
&&& \lambda_{l-1} & b_{l-1}\\
0&&&&\lambda_l
\end{array}
\right] \ .
\eeq
Here $\lambda_i$ are the eigenvalues of $A$,
and $b_i$ are
either 0 or 1 depending on the degrees of the 
elementary divisors. 
 We find that
\bey{9}
\lim_{\lambda \to 1^+} \left[ (\lambda -1)(\lambda I -A)^{-1} \right] &=&
\lim_{\lambda \to 1^+} \left[W (\lambda -1)(\lambda I -J)^{-1} W^{-1}\right] 
\nonumber \\
&=&
W \left[\lim_{\lambda \to 1^+}
  \left[ (\lambda -1)(\lambda I -J)^{-1}\right] \right] W^{-1} \ .
\eey
Carrying out the inversion, we find
\beq{10}
(\lambda I  - J)^{-1} = 
\left[
\begin{array}{ccccc}
\frac{1}{\lambda -1}&
\frac{b_1}{(\lambda -1)(\lambda -\lambda_2)}&
\frac{b_1 b_2}{(\lambda -1)(\lambda -\lambda_2)(\lambda -\lambda_3)}&
\frac{b_1 b_2 b_3}{(\lambda -1)(\lambda -\lambda_2)(\lambda -\lambda_3)(\lambda -\lambda_4)}&
\ldots\\
0 &  \frac{1}{\lambda -\lambda_2}&
\frac{b_2}{(\lambda -\lambda_2)(\lambda -\lambda_3)}&
\frac{b_2 b_3}{(\lambda -\lambda_2)(\lambda -\lambda_3)(\lambda -\lambda_4)}&
\ldots\\
0 & 0& \frac{1}{\lambda -\lambda_3}&
\frac{b_3}{(\lambda -\lambda_3)(\lambda -\lambda_4)}&
\ldots \\
0&0&0& \frac{1}{\lambda -\lambda_4}&
\ldots\\
\vdots & \vdots & \vdots & \vdots&\ \hspace{-1.0 in} \ddots\\
0&0&0&0& \frac{1}{\lambda -\lambda_l}
\end{array}
\right] \ .
\eeq
Now since $\lambda_1 = 1$ is the only eigenvalue of unit modulus,
the associated elementary divisor has degree 1, and $b_1 = 0$.
We then see that
\beq{11}
\lim_{\lambda \to 1^+} \left[ (\lambda -1)(\lambda I -J)^{-1}  \right]
=
\left[
\begin{array}{ccccc}
1 &0&\ldots\\
0&0&\ldots\\
\vdots & \vdots & \ddots
\end{array}
\right] = {\bf e} {\bf e}^\top \ ,
\eeq
where $e_1 = 1$, and $e_i = 0$ otherwise.
We denote the columns of $W$ by ${\bf w}^{(i)}$.  Since 
$W$ is by definition a  non-singular matrix,\cite{Grantmacher}
the vectors
${\bf w}^{(i)}$ form a complete basis,
and we can decompose the
initial condition as ${\bf x}_0 = \sum_{i=1}^l a_i {\bf w}^{(i)} 
\equiv W {\bf a}$.  From Eqs.\ (\ref{7}), (\ref{9}), and (\ref{11}), we have
\bey{12}
\lim_{n \to \infty}A^n {\bf x}_0 &=& \lim_{\lambda \to 1^+}
\left[ (\lambda -1)(\lambda I - A)^{-1} \right] {\bf x}_0 
\nonumber \\
&=& W {\bf e} {\bf e}^\top W^{-1} W {\bf a} 
\nonumber \\
&=& W {\bf e} {\bf e}^\top {\bf a} 
\nonumber \\
&=& a_1 {\bf w}^{(1)}
\nonumber \\
&=& {\bf w}^{(1)} \ ,
\eey
where in the last step we have used Eq.\ (\ref{3}) and the fact that 
both ${\bf x}_0$ and ${\bf w}^{(1)}$ are normalized
probability distributions.
This completes the rigorous proof of the theorem.

The implications of our theorem are significant.  First,
the sequential updating approach, widely practiced and nearly as
widely criticized, is correct.  All that is required is
that the sweep leave the Boltzmann distribution invariant, in 
addition to leading to regular sampling.
Indeed, there are numerical
reports that the sequential updating approach
leads not only to correct sampling but also 
to \emph{more efficient} sampling in a variety of contexts, ranging
from image processing to membrane physics.\cite{Marco}
There are some simple cases, however, for which
random Metropolis updating
is ergodic and regular, but sequential Metropolis updating is not.  One is the
Ising model at infinite temperature, where the
configurations produced by sequential updating
simply oscillate between the initial configuration and 
the configuration with all spins flipped.  Another example is the
$2 \times 2$ Ising model at any temperature.  If the 
spins are updated in the order 
$\left[ \begin{array}{cc} 1 & 2 \\ 3 & 4 \end{array} \right] $,
then the initial configurations
$\left[ \begin{array}{cc} + & - \\ + & + \end{array} \right] $ and
$\left[ \begin{array}{cc} + & + \\ - & + \end{array} \right] $
lead to sequences that also
oscillate between spin-flipped configurations.
In these examples, our theorem does not guarantee
convergence to the Boltzmann distribution,
because assumption (\ref{4}) is violated.
If, however, the  individual spin-flip
 moves within the sweep were performed with a
probability $1-\alpha$, as in Eq.\ (\ref{4a}), then 
convergence to the Boltzmann distribution would occur in both
of these examples.

Note that our theorem requires the sampling to be \emph{regular}, not
simply ergodic.  Indeed, a method that is ergodic 
may not converge to the invariant distribution, even if
it satisfies detailed balance.
A simple example is provided by a two-state system.  The transition
matrix
$\left[ \begin{array}{cc} 0 & 1 \\ 1 & 0 \end{array} \right] $
leads to ergodic, but not regular, sampling.  This transition
matrix  satisfies
detailed balance with the invariant distribution $x^*_1=x^*_2 = 1/2$.
The limit $\lim_{n \to \infty} A^n x_0$ does not exist for a
general initial condition, however, 
because $A^n x_0$ simply alternates between $x_0$ and the swapped
state for even and odd $n$.

There has been concern that perhaps one should not use the data
obtained from within the sweep of a sequential move, or that perhaps 
the statistics collected would depend on how the sweep is defined.
In fact, we can show that all of the data can be used, and the
results do not depend on how the sweep is defined.  For simplicity,
let us consider a sweep composed of $N$ steps: $A = \prod_{k=1}^N
B^{(k)}$.  Our general theorem says that the data collected via the
procedure
${\bf x}_n = A^n {\bf x}_0$ will sample the Boltzmann distribution.
Moreover, after the initial equilibration period, the initial condition
${\bf x}_0$ will not affect the results.  We can, therefore, equally
well consider the exact same sequence of moves but
collect data at a shifted period within the sweep: ${\bf x}_n = 
[B^{(N)} \prod_{k=1}^{N-1} B^{(k)} ]^n [B^{(N)} {\bf x}_0]$.  This approach,
which uses the data within the sweep, is also susceptible
to our general theorem.  We can generalize this argument
to any arbitrary  shifting of the data collection period.
By our theorem,
all of the data so collected
will be sampling from the Boltzmann distribution once the
equilibration regime has passed.
The data from  within the sweep, therefore, can be used, and the
average statistics will not depend on which data are used.

We note that a simple way to ensure that a sequential sweep
satisfy the balance condition is to require that each of the
moves within the sweep satisfy local detailed balance:
$B^{(k)}_{ij} x^*_j = B^{(k)}_{ji} x^*_i$. 
Indeed, the local detailed balance
condition is satisfied in all of the sequential Monte Carlo methods
previously cited.
 By using this
condition, one can show that the balance condition,
Eq.\ (\ref{5}), is satisfied:
\bey{13}
[A {\bf x}^*]_i &= & [\prod_{k=1}^N B^{(k)} {\bf x}^*]_i
\nonumber \\ &=& 
 \sum_{i_1, i_2, \ldots, i_N} 
B_{i i_1}^{(1)}
B_{i_1 i_2}^{(2)} \cdots
B_{i_{N-2} i_{N-1}}^{(N-1)}
B_{i_{N-1} i_N}^{(N)} x_{i_N}^*
\nonumber \\ &=& 
 \sum_{i_1, i_2, \ldots, i_N} 
B_{i i_1}^{(1)}
B_{i_1 i_2}^{(2)} \cdots
B_{i_{N-2} i_{N-1}}^{(N-1)} x_{i_{N-1}}^*
B_{i_{N} i_{N-1}}^{(N)}
\nonumber \\ &=& 
 \sum_{i_1, i_2, \ldots, i_{N-1}} 
B_{i i_1}^{(1)}
B_{i_1 i_2}^{(2)} \cdots
B_{i_{N-2} i_{N-1}}^{(N-1)} x_{i_{N-1}}^*
\nonumber \\
&\vdots&
\nonumber \\
&=& x_i^* \ .
\eey
Here we have used local detailed balance condition to achieve the third line
and the fact that the transpose of each $B^{(k)}$ is stochastic to
achieve the fourth line.  The final line is achieved by induction.

More generally, our result implies that any Monte Carlo method
that leaves the Boltzmann distribution invariant is correct as long
as it leads to regular, Markov sampling.
The type-R and type-M transitions within the
dynamic weighting method of Wong and Liang, for example, 
are known only to leave the Boltzmann distribution invariant.\cite{Wing}
Although the method led to efficient and apparently accurate
sampling, the formal correctness of this method has not been shown.
Our general theorem formally establishes the validity of this and other
such methods.

In summary, we have clarified a long-standing and controversial
issue in Monte Carlo simulations by establishing the balance condition
as a necessary and sufficient
fundamental requirement.  This condition is substantially weaker
than the detailed balance condition.  Although our proof of
convergence demands that
the set of Monte Carlo moves lead to regular sampling, this is the
case for almost all sets of moves that lead to ergodic sampling.
We have shown that the local detailed balance condition is one means
of constructing a sequential Monte Carlo scheme satisfying the balance
condition.

\section*{Acknowledgment}
This research was supported by the National Science Foundation.
It is a pleasure to acknowledge stimulating discussions with
Daan Frenkel, Giorgio Parisi, and Alan D.\ Sokal.
\bigskip
%%%%%%%%%%%%%%%%%%%%%%%%%%%%%%%%%%%%%%%%%%%%%%%%%%%%%%%%%%%%%%%%%%%%%%%%%%%%

\bibliography{balance}

\end{document}